\documentclass[letterpaper, 10 pt, conference]{ieeeconf}  

\IEEEoverridecommandlockouts                              

\usepackage{graphics} 
\usepackage{epsfig} 
\usepackage{mathptmx} 
\usepackage{times} 
\usepackage{amsmath} 
\usepackage{amssymb}  
\usepackage{physics}
\usepackage{mathtools}
\usepackage{float}
\usepackage{caption}
\usepackage{subcaption}
\usepackage{algorithm}
\usepackage{algpseudocode}
\usepackage{xcolor}
\usepackage[]{mdframed}
\usepackage{fancyhdr}

\title{\LARGE \bf
Weighted Feedback-Based Quantum Algorithm for Excited States
Calculation}

\author{ Salahuddin Abdul Rahman$^{*}$, Özkan Karabacak$^{\dagger}$, Rafal Wisniewski$^{*}$\\\\
        *Section of Automation and Control, Department of Electronic Systems, Aalborg University, Aalborg, Denmark\\
        $\dagger$Department of Mechatronics Engineering, Kadir Has University, Istanbul, Turkey
}

\fancypagestyle{firstpage}{
  \fancyhf{} 
  \fancyhead[C]{Accepted for  Publication in 2024 IEEE International Conference on Quantum Computing and Engineering (QCE)}
}

\begin{document}

\maketitle
\thispagestyle{empty}
\pagestyle{empty}

\thispagestyle{firstpage}
\begin{abstract}
Drawing inspiration from the Lyapunov control technique for quantum systems, feedback-based quantum algorithms have been proposed for calculating the ground states of Hamiltonians. In this work, we consider extending these algorithms to tackle calculating excited states. Inspired by the weighted subspace-search variational quantum eigensolver algorithm, we propose a novel weighted feedback-based quantum algorithm for excited state calculation. We show that depending on how we design the weights and the feedback law, we can prepare the $p$th excited state or lowest energy states up to the $p$th excited state. Through an application in quantum chemistry, we show the effectiveness of the proposed algorithm, evaluating its efficacy via numerical simulations.
\end{abstract}

\section{Introduction} \label{s1}
One of the primary applications where noisy intermediate-scale quantum (NISQ) devices are expected to exhibit an advantage over classical algorithms is preparing eigenstates of Hamiltonians. Variational quantum algorithms (VQAs) are considered the leading algorithms for NISQ devices, fulfilling their requirements and anticipated to demonstrate quantum advantage.  For an in-depth exploration of VQAs and their applications, see the comprehensive review by Cerezo et al. \cite{cerezo2021variational}. Nevertheless, the effectiveness of VQAs encounters challenges, such as the complex structuring of ansatz designs and the requirement to address computationally intensive classical optimization tasks for updating the circuit parameters. Proposals to overcome these challenges were proposed, such as dynamical construction of the parameterized quantum circuit \cite{tang2021qubit,zhu2022adaptive} and employing efficient classical optimizers \cite{lavrijsen2020classical}.  

A proposed alternative to VQAs for addressing quadratic unconstrained binary optimization (QUBO) problems is the feedback-based algorithm for quantum optimization (FALQON) introduced by Magann et al. in \cite{magann2022feedback,magann2022lyapunov}. FALQON has the advantage of dynamically constructing the quantum circuit and avoiding classical optimization to update the circuit parameters. Instead, it assigns the circuit parameters by measuring the qubits from the preceding layer and using a feedback control law based on the quantum Lyapunov control (QLC) theory. FALQON has also been applied to solve constrained optimization problems \cite{DavidWakeham2021}. Furthermore, as explored in \cite{magann2022lyapunov}, FALQON can serve as an initialization technique for the parameters of the quantum approximate optimization algorithm (QAOA), enhancing its efficiency. This application renders FALQON suitable for NISQ devices. In \cite{larsen2023feedback}, the feedback-based quantum algorithm (FQA) was introduced as a generalization of FALQON for finding ground states of systems of interacting electrons. In \cite{magann2023randomized}, a randomized adaptive approach for preparing quantum states is introduced that can be combined with feedback-based quantum algorithms to improve the convergence properties of such algorithms.

Calculating excited states of Hamiltonians has applications in quantum chemistry \cite{motta2022emerging} and combinatorial optimization \cite{ura2023analysis}. Various quantum algorithms have been proposed in the literature to tackle the problem of calculating excited states, including quantum imaginary-time evolution (QITE) algorithms \cite{motta2020determining}, quantum annealing (QA) algorithms \cite{seki2021excited}, quantum phase estimation (QPE) algorithms \cite{bauman2020toward}, VQAs \cite{cerezo2021variational}, and others \cite{wen2024full,cortes2022quantum}.

For VQAs, several approaches have been proposed to extend the variational quantum eigensolver (VQE) framework to calculate excited states, such as the subspace-search variational quantum eigensolver for excited states (SSVQE) \cite{nakanishi2019subspace}, the variational quantum deflation (VQD) \cite{higgott2019variational}, quantum subspace expansion (QSE) \cite{mcclean2017hybrid}, constrained VQE \cite{ryabinkin2018constrained} and others \cite{santagati2018witnessing}. VQD and SSVQE utilize the fact that the eigenstates of any given Hamiltonian are orthogonal. In VQD, an overlap term is added in the cost function to impose orthogonality with all lower energy eigenstates. However, this technique requires knowledge of all lower energy eigenstates and estimating overlap terms to evaluate the cost function. Weighted SSVQE algorithm exploits the fact that unitary transformation preserves orthogonality. Hence, starting from initial orthogonal states and under unitary transformation, orthogonality is preserved at output states. SSVQE can find up to the $p$th excited state using one optimization procedure.

In \cite{rahman2024feedback}, the feedback-based quantum algorithm for excited states calculation (FQAE) was introduced as an extension of FQAs to calculate excited states of Hamiltonians utilizing tools from QLC theory and deflation techniques \cite{mackey2008deflation}. To prepare the $p$th excited state, FQAE assumes knowledge of all the lower energy eigenstates. Therefore, to prepare the $p$th excited state, FQAE should be applied iteratively first to generate all the lower energy eigenstates. Inspired by weighted SSVQE, this work proposes the weighted feedback-based quantum algorithm for excited states calculation (WFQAE). WFQAE can calculate the lowest energy states up to the $p$th excited state using a single optimization procedure.

The subsequent sections of this paper are organized as follows: Section~\ref{s2} provides an overview of FALQON and its relation to QLC. In Section~\ref{s3}, we present WFQAE. Section~\ref{s4} explores the algorithm's efficiency through application in quantum chemistry. Lastly, Section~\ref{s5} provides a conclusion and future work.

\section{Preliminary} \label{s2}
In this section, we give a general overview of FALQON and its relation to QLC \cite{magann2022feedback,magann2022lyapunov}.

Consider the Hilbert space $\mathcal{H}=\mathbb{C}^L$ equipped with the associated orthonormal basis $\mathcal{E}=\{|j\rangle\}_{j \in\{0, \ldots, L-1\}}$ and the set of quantum states denoted by $\mathcal{X}=\{\ket{\phi} \in \mathbb{C}^L : \braket{\phi}=\| \ket{\phi} \|^2=1 \}$. Hereafter, all operators will be expressed in terms of the $\mathcal{E}$ basis.
Let us consider a quantum system governed by the controlled time-dependent Schrödinger equation:

\begin{equation}
    i\ket{\dot{\phi}(t)}=(H_d+\alpha(t)H_c)\ket{\phi(t)},
    \label{model1}
\end{equation}
where $\hslash$ is normalized to $1$, $\alpha(t)$ denotes the control input, $H_d$ is the drift Hamiltonian with associated eigenvaules $E_0< E_1< \cdots <E_{L-1}$ and corresponding eigenvectors $\ket{E_0},\ket{E_1},\dots,\ket{E_{L-1}}$, and $H_c$ is the control Hamiltonian. 
Throughout this work, we assume that the Hamiltonians $H_d$ and $H_c$ are non-commuting and time-independent, implying $[H_d, H_c]\neq0$. 

The primary objective here is to devise a feedback control law, $\alpha(\ket{\phi(t)})$, ensuring the convergence of the quantum system described in \eqref{model1} to the ground state of the Hamiltonian $H_d$, for all initial states. In other words, we seek to attain the state $\ket{\phi_g}=\text{argmin}_{\ket{\phi}\in \mathcal{H}} \bra{\phi}H_d\ket{\phi}$.

Let $ V(\ket{\phi}) = \bra{\phi}H_d\ket{\phi}$ be the designed Lyapunov function, where its derivative along the trajectories of system \eqref{model1} is given by $\dot{V}(\ket{\phi(t)}) =  \bra{\phi(t)}  \mathrm{i}[H_c,H_d] \ket{\phi(t)} \alpha(t)$. Consequently, by designing $\alpha(t)$ according to 
\begin{equation}
    \alpha(t)= - \bra{\phi(t)}  \mathrm{i}[H_c,H_d \ket{\phi(t)},
    \label{controller1}
\end{equation}
it ensures that $\dot{V}\leq0$. The application of the controller \eqref{controller1}, under certain assumptions (see Appendix A of \cite{magann2022lyapunov}), guarantees asymptotic convergence to the ground state $\ket{\phi_g}$, for almost all initial states.

The solution of \eqref{model1} is given as $ U(t)=\eta e^{-i \int_{0}^{t} H(r) d r}$, where $\eta$ is the time-ordering operator. By dividing it into $l$ intervals of constant duration $\Delta t$, we approximate $U(T, 0)$ as $\prod_{k=1}^{l} e^{-i H(k \Delta t) \Delta t}$, with $\Delta t$ chosen small enough for $\mathrm{H}(t)$ to remain approximately constant within each interval. Employing Trotterization simplifies this to $U(T, 0) \approx \prod_{k=1}^{l} e^{-i \alpha(k \Delta t) H_{c}\Delta t} e^{-iH_{d}\Delta t}$. Consequently, we obtain a digitized representation of the evolution in the form
\begin{align}
\ket{\phi_l}   &=\prod_{k=1}^{l} \big(e^{-i \alpha(k \Delta t)  H_{c}\Delta t}e^{-iH_{d}\Delta t} \big) \ket{\phi_0} =  \prod_{k=1}^{l}  \big( U_c(\alpha_k)U_d \big ) \ket{\phi_0} ,
\label{evolution1}
\end{align}
where the following notation is adopted. $\alpha_k=\alpha (k \Delta t)$, $U_c(\alpha_k)=e^{-i \alpha(k \Delta t)  H_{c}\Delta t}$, $\ket{\phi_k}=\ket{\phi(k \Delta t)}$, $U_d=e^{-iH_{d}\Delta t}$. Note that the values of the controller at each discrete time step represent the circuit parameters. Throughout this work, the terms \emph{controller} and \emph{circuit parameters} will be used interchangeably. 

We assume that the drift Hamiltonian $H_d$ is given as a sum of Pauli strings in the form $H_d=\sum_{j=1}^{N_0} c_j S_j$, where each Pauli string $S_j$ is a Hermitian operator represented as $S_j=S_{j,1} \otimes S_{j,2} \otimes ... \otimes S_{j,n}$ with $S_{j,n} \in \{I,\sigma_x,\sigma_y,\sigma_z \}$, and $c_j$ represents real scalar coefficients while $N_0$ is a polynomial function of the qubit count. Consequently, the unitary transformation $U_d$ can be efficiently implemented as a quantum circuit. Likewise, for effective implementation of the operator $U_c$ as a quantum circuit, the control Hamiltonian $H_c$ should be designed as $ H_c = \sum_{j=1}^{N_1} \bar c_j \bar S_j $. For details on the implementation of $U_d$ and $U_c(\alpha_k)$ see \cite{magann2021digital}.

The quantum circuit that implements $U(T,0)$ simulates the propagator $U(t)$, where choosing $\Delta t$ sufficiently small can guarantee that $\dot V\leq 0$ \cite{magann2022lyapunov}. For the feedback law, the following discrete version of the controller \eqref{controller1} is adopted:
\begin{align} \label{controller1_dis}
    \alpha_{k+1} &= - \bra{\phi_k} \mathrm{i} [H_c,H_d]  \ket{\phi_k}.   
\end{align}

The algorithm is initialized by assigning an initial value to $\alpha_1=\alpha_\text{init}$ and specifying the time step $\Delta t$. Next, a set of qubits is initialized to an easily preparable state $\ket{\phi_0}$, followed by applying one circuit layer to prepare the state $\ket{\phi_1}$. To estimate the controller for the next layer $\alpha_2$, the controller 
is expanded in terms of Pauli strings in the following way:
\begin{align} \label{Pau}
    \alpha_{k+1} &= \sum_{j=1}^{N_2} \hat c_j \bra{\phi_k} \hat S_j \ket{\phi_k},
\end{align}
where $\hat S_j$ represents a Pauli string, $N_2$ denotes the total number of Pauli strings. Note that $N_2$ depends on both $N_0$ and $N_1$, and given that $N_0$ and $N_1$ are polynomial functions of the qubit count, $N_2$  is similarly a polynomial function of the qubit count.  Hence, the controller can be evaluated by estimating each expectation in \eqref{Pau}. Next, the procedure of adding a new layer and estimating the controller for the next layer is iteratively repeated for $l$ layers. The resultant quantum circuit, denoted as $\prod_{k=1}^{l} \big(U_c(\alpha_k)U_d\big)$ along its parameters $\{\alpha_k\}_{k=1,\dots,l}$ serve as the output of the algorithm. This output approximates the ground state of the Hamiltonian $H_d$. 

\algnewcommand\algorithmicInput{\textbf{Input:}}
\algnewcommand\algorithmicOutput{\textbf{Output:}}
\algnewcommand\Input{\item[\algorithmicInput]}
\algnewcommand\Output{\item[\algorithmicOutput]}
   
    \begin{algorithm} [H]
    \caption{FALQON \cite{magann2022lyapunov}}\label{FALQON}
    \begin{algorithmic}[1]
    \Input{$H_d$, $H_c$, $\Delta t$, $p$, $\ket{\phi_0}$}
    \Output{The quantum circuit to prepare the ground state along its parameters $\{\alpha_k\}_{k=1,\dots,l}$}
    \State{Set $\alpha_1=0$}
    \State{\textbf{Repeat} at every step $k = 1, 2, 3, \dots, l-1$}
            \State{Prepare the initial state $\ket{\phi_0}$}
            \State {Prepare the state $\ket{\phi_k}=\prod_{l=1}^{k} \big(U_c(\alpha_k)U_d \big)\ket{\phi_0} $}
            \State {Calculate the controller $\alpha_{k+1}$ by estimating each expectation term in \eqref{Pau} }
            \State {\textbf{Until $k=l$}}

    \end{algorithmic}
    \end{algorithm}

\section{Weighted Feedback-based Quantum Algorithm for Excited States Calculation} \label{s3}
This section introduces the problem of calculating
excited states of a given Hamiltonian and suggests a solution using QLC. Based on this solution, we propose WFQAE.

Consider the following model:
\begin{align} \label{augmented}
\mathrm{i}\ket*{\dot\phi^{(0)}(t)}  &=(H_d+\alpha(t)H_c)\ket*{\phi^{(0)}(t)}, \nonumber \\
\mathrm{i}\ket*{\dot\phi^{(1)}(t)} &=(H_d+ \alpha(t)H_c)\ket*{\phi^{(1)}(t)}, \nonumber \\
    &\quad \quad \quad \vdots \nonumber  \\
\mathrm{i}\ket*{\dot\phi^{(p)}(t)} &=(H_d+\alpha(t)H_c)\ket*{\phi^{(p)}(t)}, 
\end{align}
where the initial states $\{ \ket*{\phi_0^{(q)} } \}_{q=0,\dots,p}$ are assumed to be orthogonal (i.e. $\braket*{\phi_0^{(q)}}{\phi_0^{(j)}}=\delta_{q,j}$). For the simplicity of analysis, we consider one control input $\alpha(t)$. The analysis can easily be extended to encompass multiple control inputs, as further explored in the appendix. Note that since the eigenstates of Hermitian operators are orthogonal, and since we want the states to evolve into the eigenstates of $H_d$ eventually, we choose the initial states to be orthogonal.

Consider the state of the composite system:
\begin{equation}
    \ket{\Phi} = \ket*{\phi^{(0)}} \otimes \ket*{\phi^{(1)}} \otimes \dots \otimes \ket*{\phi^{(p)}}.
\end{equation}
In addition, consider the following Hamiltonians:
\begin{equation}
    \hat H_d = \sum_{q=0}^{p} H_d^{(q)},
\end{equation}
where we adopt the notation $H_d^{(q)} = I \otimes \dots \otimes \underbrace{H_d}_{q\text{th position}} \otimes \dots I$. Similarly, we have 
\begin{equation}
    \hat H_c = \sum_{q=0}^{p} H_c^{(q)}.
\end{equation}
Since $H_d^{(q)}$ and $H_d^{(j)}$ commute for any $q$ and $j$ in $\{0, \hdots, p\}$, the model in \eqref{augmented} is equivalent to the following model:
\begin{equation}
    \mathrm{i} |\dot{\Phi}(t)\rangle=(\hat H_d+\alpha(t)\hat H_c)|\Phi(t)\rangle,
    \label{model2}
\end{equation}
with the initial state defined as 
\begin{equation}
    \ket{\Phi_0} = \ket*{\phi_0^{(0)}} \otimes \ket*{\phi_0^{(1)}} \otimes \dots \otimes \ket*{\phi_0^{(p)}}.
\end{equation}
Define the set $\mathcal{Y}=\{\ket{\Phi} \in \left(\mathbb{C}^L\right)^{\otimes p+1} : \braket{\Phi}=\| \ket{\Phi} \|^2=1 \}$.
Since $\ket{\Phi_0}$ is separable and the composite system in \eqref{model2} is decoupled, the solutions remain separable. Additionally, since the initial states $\{ \ket*{\phi_0^{(q)} } \}_{q=0,\dots,p} $ are orthogonal and evolution is unitary, orthogonality is also preserved. As a result, solutions of \eqref{model2} remain in a subset $\bar{\mathcal Y}\subset \mathcal Y$, defined as the set of (separable) states that can be written as a tensor product of $p$ orthogonal vectors in $\mathcal H$.

Let $Q$ be a Hermitian operator defined as follows:
\begin{equation}   
Q := \sum_{q=0}^p w_q H_d^{(q)},
\end{equation} 
where the weights $\{ w_q \}_{q=0,\dots,p} $ are chosen such that $w_q>w_j$ for $q<j$.
Consider a Lyapunov function defined in the set $\bar{\mathcal Y}$ in the following form:
\begin{align} \label{WL}
    V&=\bra{\Phi} Q \ket{\Phi}, \nonumber \\
     &= \sum_{q=0}^p w_q \bra*{\phi^{(q)}(t)}H_d\ket*{\phi^{(q)}(t)}.
\end{align}
The minimum of $V$ on $\bar{\mathcal Y}$ is then attained at the desired state $(\ket{E}=\ket{E_0}\otimes\ket{E_1} \otimes \cdots \otimes \ket{E_p})$, which is the minimum energy state of $Q$ in the set $\bar{\mathcal Y}$  (see appendix of \cite{nakanishi2019subspace}). In this case, the objective is to design a feedback control law, $\alpha(\ket{\Phi(t)})$, ensuring the convergence of the quantum system described in \eqref{model2} to the state $\ket{E}$, for all initial states.

The derivative of $V$ along the trajectories of system \eqref{model2} is given by:
\begin{align}
\dot{V} & = \sum_{q=0}^p w_q \bra*{\phi^{(q)}(t)} \mathrm{i} [H_c,H_d] \ket*{\phi^{(q)}(t)} \alpha (t).
\label{vdot2}
\end{align}
We design $\alpha (t)$ such that $\dot{V}\leq0$:
\begin{equation} \label{controller2}
    \alpha (t)= - K_g   h \Big( \sum_{q=0}^p w_q \bra*{\phi^{(q)}(t)} \mathrm{i} [H_c,H_d] \ket*{\phi^{(q)}(t)} \Big),
\end{equation}
where $K_g>0$ and $h$ is a continuous function satisfying $h(0) = 0$ and $xh(x) >
0$, for all $x \neq 0$. 

The application of the controller \eqref{controller2}, under certain assumptions (see Appendix A of \cite{magann2022lyapunov}), guarantees asymptotic convergence to the state  $\ket{E}$ for almost all initial states \cite{cong2013survey,magann2022lyapunov}.

Utilizing a similar Trotterization procedure as presented in Section~\ref{s2}, we get the following:
\begin{align}
\ket*{\phi^{(q)}_l}   &=\prod_{k=1}^{l} \bigl( e^{-i \alpha(k \Delta t)  H_{c}\Delta t}e^{-iH_{d}\Delta t} \bigr) \ket*{\phi_0^{(q)}}, \nonumber \\
&=  \prod_{k=1}^{l} \bigl(U_c(\alpha_k)U_d \bigr) \ket*{\phi_0^{(q)}}.
\label{evolution2}
\end{align}
For the controller, the following discrete form of the feedback law \eqref{controller2} is adopted:
\begin{align} \label{controller2_dis}
    \alpha_{k+1} &= - K_g   h \Big( \sum_{q=0}^p w_q \bra*{\phi^{(q)}_k} \mathrm{i} [H_c,H_d] \ket*{\phi^{(q)}_k} \Big), \nonumber \\
    &= - K_g    \sum_{q=0}^p w_q B_k^{(q)}  ,       
\end{align}
where $B_k^{(q)} = \bra*{\phi^{(q)}_k} \mathrm{i} [H_c,H_d] \ket*{\phi^{(q)}_k}$ and $h(\cdot)$ is chosen to be the identity function.

\begin{figure}[H]
    \centering
    \includegraphics[ width=1 \linewidth]{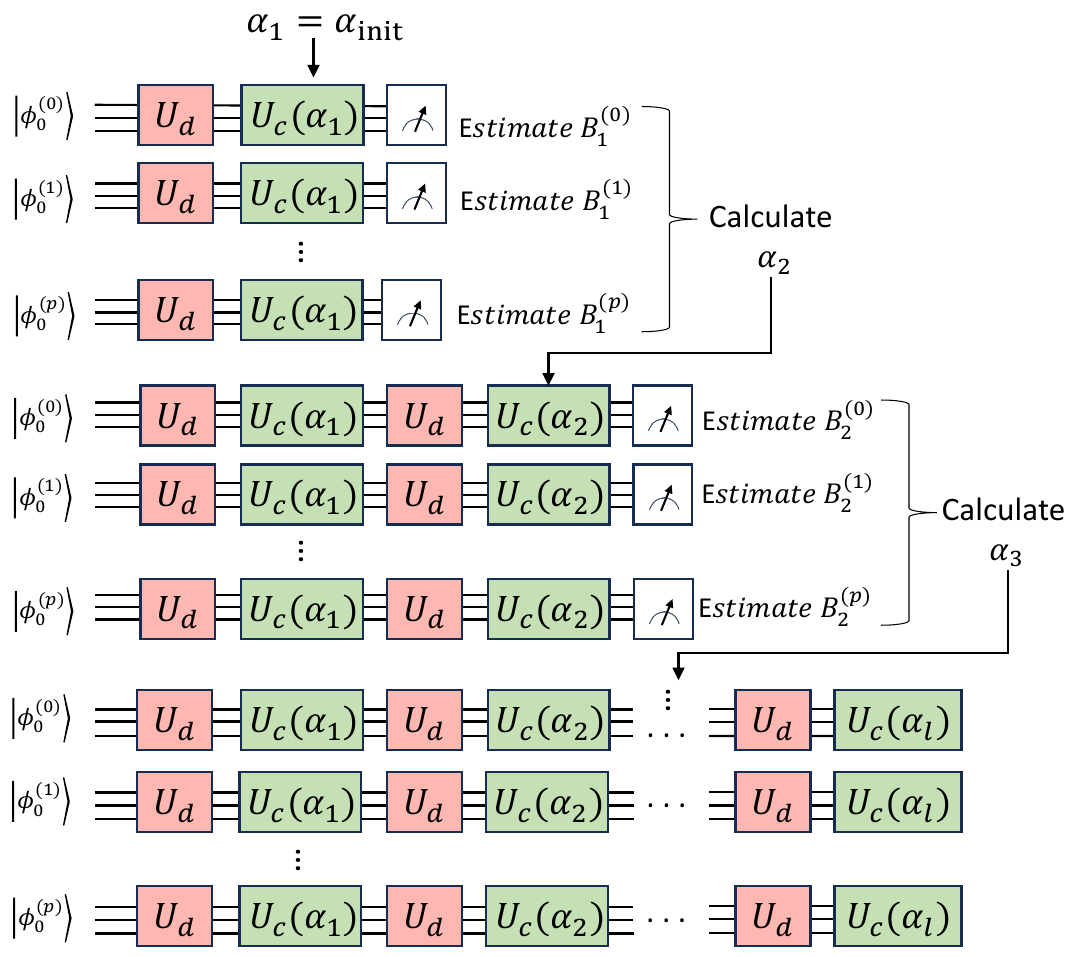}
    \caption{This figure represents the algorithmic steps for WFQAE. The algorithm starts by setting an initial value for the circuit parameter of the first layer $\alpha_1 = \alpha_{\text{init}}$. Subsequently, for all $q \in \{0,1, \dots, p \}$, the state $\ket*{\phi^{(q)}}$ is prepared by applying one circuit layer to the initial state $\ket*{\phi^{(q)}} = U_c(\alpha_k)U_d \ket*{\phi_0^{(q)}}$. The qubits are then measured to estimate the term $B_k^{(q)}$. The circuit parameter for the next layer $\alpha_2$ is then calculated either using \eqref{controller3_dis} if the target is only the $p$th excited state or using \eqref{controller2_dis} if the target is to prepare lowest energy eigenstates up to the $p$th excited state. 
    Subsequently, for the ${l-1}$ iterations, the quantum circuit is expanded by appending a new layer of $U_c(\alpha_k)U_d$, where the controller $\alpha_k$ is calculated according to \eqref{controller2_dis} or \eqref{controller3_dis}. Upon reaching the maximum depth $l-1$, the procedure concludes, and the algorithm's output is the designed quantum circuit along its parameters $\{\alpha_k\}_{k=1,\dots,l}$.}
    \label{FQAalg}
\end{figure}

The algorithmic steps for WFQAE to prepare the lowest energy states up to the $p$th excited state are detailed in Algorithm~2. The initial step involves seeding the procedure with an initial value for the parameter of the first quantum circuit layer $\alpha_1=\alpha_{\text{init}}$ and setting a value for the time step $\Delta t$. Subsequently, for all $q \in \{0,1, \dots, p\}$, the state $\ket*{\phi^{(q)}_k}$ is prepared by applying one layer of the quantum circuit to an initial state $\ket*{\phi^{(q)}_0}$. Note that besides being mutually orthogonal, the initial states $\{ \ket*{\phi_0^{(q)} } \}_{j=0,\dots,p} $ are chosen to be easily preparable on a quantum computer, thereby facilitating the implementation of the algorithm. The qubits are then measured to estimate each of $B_k^{(q)}$. To do so, we expand this term in a Pauli basis as follows:
\begin{equation}
    \label{expectation}
     B_k^{(q)}=     \bra*{\phi^{(q)}_k} \mathrm{i} [H_c,H_d] \ket*{\phi^{(q)}_k}
     = \sum_{j=1}^{N_3} \hat{c}_j \bra*{\phi^{(q)}_k} \hat{S}_j\ket*{\phi^{(q)}_k},
\end{equation}
where $\hat{S}_j$ is a Pauli string, $N_3$ is the number of Pauli strings and $\hat{c}_j$ is a real coefficient. Each of the expectations $\bra*{\phi^{(q)}_k} \hat{S}_j\ket*{\phi^{(q)}_k}$ is estimated, and their values are used to evaluate the term $B_k^{(q)}$. These values are then used to evaluate the controller for the next layer using \eqref{controller2_dis}. Subsequently, an additional layer is appended to the circuit, the controller for the next layer is evaluated, and the process iterates until the desired depth of $l$ layers is achieved.
The algorithmic procedure for WFQAE is shown in Figure~\ref{FQAalg}.

    \begin{algorithm} [H]
    \caption{WFQAE to calculate the lowest energy states up to the $p$th excited state }\label{WFQAE}
    \begin{algorithmic}[1]
    \Input{$H_c$, $H_d$, $\Delta t$, $l$, orthogonal initial states $\{ \ket*{\phi_0^{(q)} } \}_{q=0,\dots,p} $, weights $\{ w_q \}_{q=0,\dots,p} $ such that $w_q>w_j$ for $q<j$}
    \Output{The quantum circuit to prepare up to the $p$th excited state along its parameters $\{\alpha_k\}_{k=1,\dots,l}$}
    \State{Set $\alpha_1=\alpha_\text{init}$}
    \State{\textbf{Repeat} at every step $k = 1, 2, 3, \dots, l-1$}
            \State{\textbf{For} $q \in \{0, 1,...,p\}$ \textbf{do}}
            \State {Prepare the state $\ket*{\phi_k^{(q)} }=\prod_{r=1}^{k} \big( U_c(\alpha_r)U_d \big) \ket*{\phi_0^{(q)}}$}
            \State {Estimate $B_k^{(q)}$ using \eqref{expectation}}
            \State {\textbf{End for}}
            \State {Calculate the controller $\alpha_{k+1}$ using \eqref{controller2_dis}}
            \State {\textbf{Until $k=l$}}

    \end{algorithmic}
    \end{algorithm}
    
\noindent
\textit{Remark 1:}
If we are only interested in the $p$th excited state, we only need to change the weights $\{ w_q \}_{q=0,\dots,p} $ , where we choose $w_p=w$ with $w \in (0,1)$, while $w_q=1$ for all $q \in \{0,1, \dots, p-1 \}$. This results in the Lyapunov function \eqref{WL} being modified in the following way:
\begin{align}
    V = w \bra*{\phi^{(p)}(t)}H_d\ket*{\phi^{(p)}(t)} + \sum_{q=0}^{p-1} \bra*{\phi^{(q)}(t)}H_d\ket*{\phi^{(q)}(t)}.
\end{align}
Accordingly, the control law \eqref{controller2} will be modified in the following way:
\begin{align}
    \alpha (t)& = - K_g   h \Big( w \bra*{\phi^{(p)}(t)}\mathrm{i} [H_c,H_d]\ket*{\phi^{(p)}(t)}, \nonumber \\
 &+ \sum_{q=0}^{p-1} \bra*{\phi^{(q)}(t)}\mathrm{i} [H_c,H_d]\ket*{\phi^{(q)}(t)} \Big),
    \label{controller3}
\end{align}
where its discrete version has the following form
\begin{align} 
    \alpha _{k+1}& = - K_g   h \Big( w \bra*{\phi^{(p)}_k}\mathrm{i} [H_c,H_d]\ket*{\phi^{(p)}_k}, \nonumber \\
 &+ \sum_{q=0}^{p-1} \bra*{\phi^{(q)}_k}\mathrm{i} [H_c,H_d]\ket*{\phi^{(q)}_k} \Big).
    \label{controller3_dis}
\end{align}
This results in a similar procedure as Algorithm~2 where the input weight is updated and the feedback law \eqref{controller3_dis} is used in Step~7.
Note that in this case, the operator $Q$ will have a degenerate spectrum. Hence, solutions will converge to a subspace spanned by the eigenstates of the operator $Q$ in $\bar{\mathcal Y}$ that correspond to the degenerate eigenvalues (see Theorem 1 in \cite{grivopoulos2003lyapunov}). However, this causes no problem for our purpose since the state $\ket*{\phi^{(p)}}$ is equal to $\ket{E_p}$ for all states in this subspace.

\noindent
\textit{Remark 2:} 
From \eqref{controller2_dis} and \eqref{controller3_dis}, it is seen that to evaluate the controller, we need to estimate $B_k^{(q)}$ for all $q \in \{0,1, \dots, p\}$ while for FALQON we only need to estimate one equivalent term. Hence, compared to FALQON \cite{magann2022feedback,magann2022lyapunov}, WFQAE has higher sampling complexity, which increases linearly with $p$.

\newpage
\section{Results and Discussion} \label{s4}
In quantum chemistry, fermionic Hamiltonians are mapped into qubit Hamiltonians using mappings such as Jordan-Wigner (JW) and Bravyi-Kitaev (BK) \cite{fedorov2022vqe}. The resulting Hamiltonian, regardless of the mapping used, can be written in general as a weighted sum of Pauli strings in the following form (see Subsection IV-A of \cite{motta2022emerging} for details): 
\begin{equation}
    H_M = \sum_q ^{N_m} r_q \hat{S}_q,
\end{equation}
where $r_q$ are real scalar coefficients and $N_M$ is the number of Pauli strings.

In this work, we consider the lithium hydride (LiH) molecule. The resulting Hamiltonian of applying BK transformation to the second quantization Hamiltonian using the STO-6G basis for the LiH system are presented as follows (for details, see Section V of \cite{hempel2018quantum}):
\begin{align}
H_M=  & g_0 I+g_1 \sigma_{z,0}+g_2\sigma_{z,1}+g_3 \sigma_{z,2}+g_4 \sigma_{z,1} \sigma_{z,0}+g_5 \sigma_{z,2} \sigma_{z,0} \nonumber \\
    & +g_6 \sigma_{z,2} \sigma_{z,1}+g_7 \sigma_{x,1} \sigma_{x,0}+g_8 \sigma_{y,1} \sigma_{y,0}+g_9 \sigma_{x,2} \sigma_{x,0} \nonumber \\
    & +g_{10} \sigma_{y,2} \sigma_{y,0}+g_{11} \sigma_{x,2} \sigma_{x,1}+g_{12} \sigma_{y,2} \sigma_{y,1},
\end{align}
where $\sigma_{z,j}$ is the Pauli $\sigma_{z}$ operator applied to the $j$th qubit, the coefficients $g_i$ are functions of the bond distance R. 

\begin{figure}[H]
    \centering
    \includegraphics[ width=1 \linewidth]{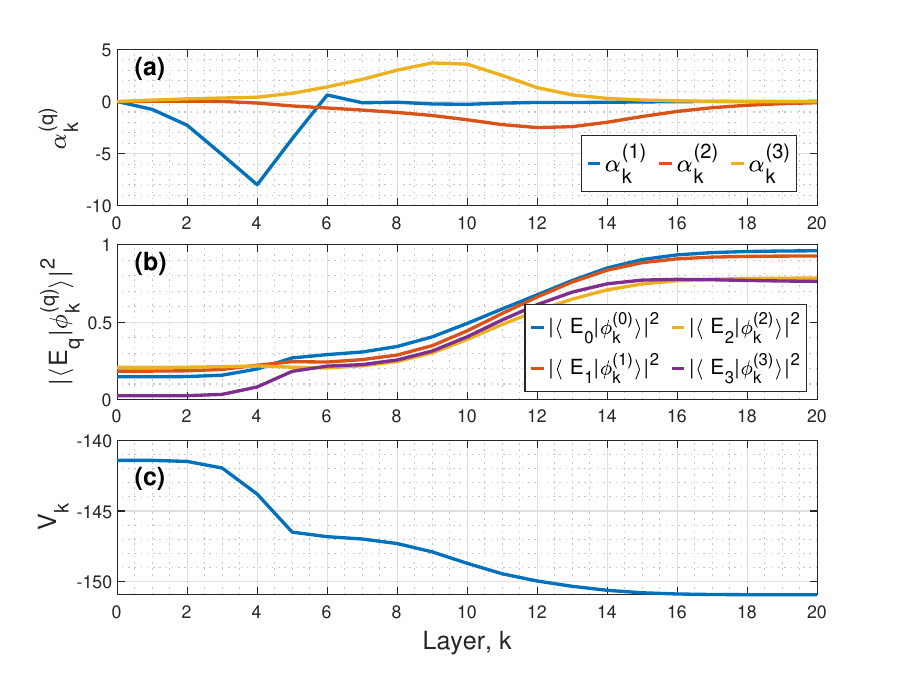}
    \caption{Simulation results for applying WFAQE to prepare up to the $3$rd excited state. The layer index k is plotted against: (a) The feedback laws $\alpha_k^{(q)}$, (b) The fidelity of the evolved states with respect to the four lowest energy eigenstates of $H_M$, and (c) The Lyapunov function $V_k = \sum_{q=0}^p w_q \bra*{\phi^{(q)}_k}H_M\ket*{\phi^{(q)}_k}$.}
    \label{SM1}
\end{figure}

\begin{figure}[H]
    \centering
    \includegraphics[ width=1 \linewidth]{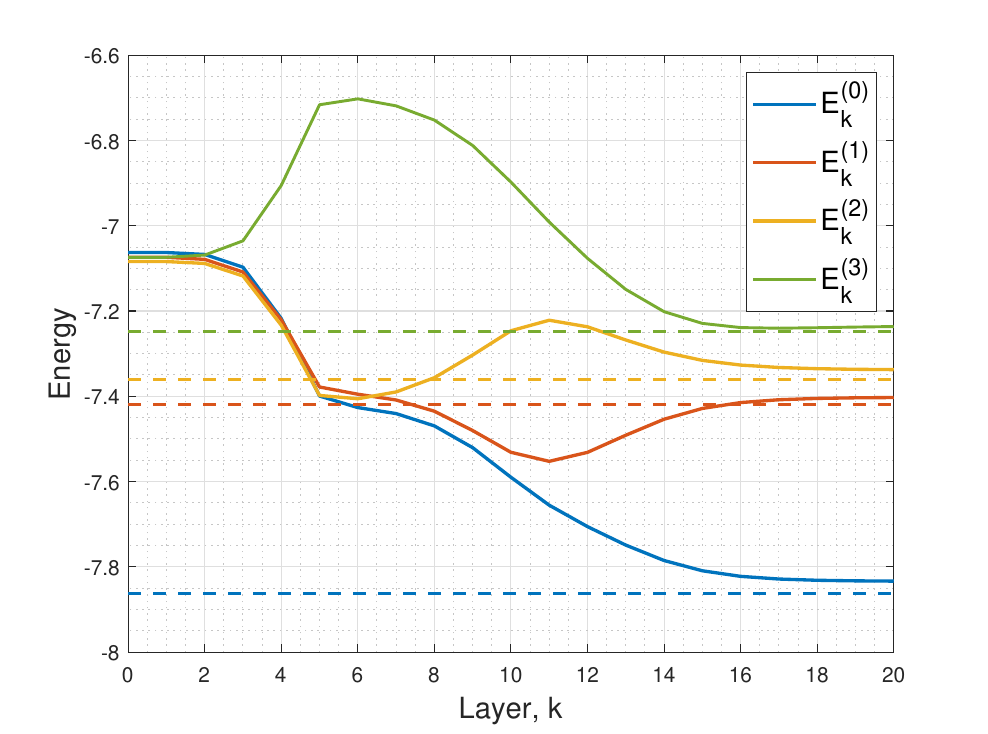}
    \caption{Simulation results for WFAQE applied to prepare up to the $3$rd excited state. The layer k is plotted versus the energy, where the energy of the $q$th excited state is defined as $E_k^{(q)}=\bra*{\phi^{(q)}_k}H_M\ket*{\phi^{(q)}_k}$.} 
    \label{SM2}
\end{figure}

The values for the numerical simulation are taken from Table 1 in \cite{zong2024determination}. For an atomic distance of $R=2.5$, we have $[g_0,g_1,g_2,g_3,g_4,g_5,g_6,g_7,g_8,g_9,g_{10},g_{11},g_{12}]=[-7.0582, 0.0094, -0.2857, -0.347, 0.0152, 0.0152, 0.0102,\\ 0.0102, 0.1957, 0.2202, 0.0208, 0.0208, 0.2563]$. We run WFQAE to calculate the four lowest energy eigenstates of the LiH molecule using the statevector simulator. We design the control Hamiltonian as  $H_c = \sum_{j=1}^3 \alpha^{(j)}H_{c,j} = \alpha^{(1)}H_{c,1}+\alpha^{(2)}H_{c,2}+ \alpha^{(3)}H_{c,3} = \alpha^{(1)} (\sigma_{z,0}+\sigma_{x,0})+\alpha^{(2)}(\sigma_{z,1}+\sigma_{x,1}) + \alpha^{(3)}(\sigma_{z,2}+\sigma_{x,2})$. We run the WFQAE for a depth of 20 layers, where the parameters are chosen as follows. We set $\Delta t = 0.05$, the weights $[w_0,w_1,w_2,w_3]=[8,6,4,2]$, the controllers' gains $[k_{g,1},k_{g,2},k_{g,3}]=[1,1,1]$ and the initial values for the controllers $[\alpha_1^{(1)},\alpha_1^{(2)},\alpha_1^{(3)}]=[0,0,0]$. The initial states are chosen as $\{ \ket*{\phi_0^{(q)} } \}_{q=0,\dots,3} =\{\ket{-++},\ket{--+},\ket{+-+},\ket{++-} \}$, known to be easily preparable and orthogonal. The simulation results are shown in Figures~\ref{SM1} and \ref{SM2}. Through numerical simulations, we note that starting from a different set of orthogonal initial states affects the convergence rate drastically. For example, for a different combination of the $\ket{+}$ and $\ket{-}$ terms we get much slower convergence rate while starting from different orthogonal initial states such as the states $\{ \ket*{\phi_0^{(q)} } \}_{q=0,\dots,3} =\{\ket{100},\ket{110},\ket{010},\ket{001} \}$ help in achieving faster convergence. Hence, valuable future work will be to investigate further how to choose suitable initial states for feedback-based quantum algorithms.

Figure~\ref{SM1} shows that the Lyapunov function monotonically decreases to its minimum, and the fidelities of the evolved states with respect to the four lowest energy eigenstates reach more than $0.75$ in $20$ layers. Figure~\ref{SM2} shows that the energies of the evolved states $E_k^{(q)}=\bra*{\phi^{(q)}_k}H_M\ket*{\phi^{(q)}_k}$ converges to the exact energies of the four lowest energy eigenstates of the Hamiltonian $H_M$.

\section{Conclusion and Future Work} \label{s5}
We have introduced WFQAE to extend the feedback-based quantum algorithms for calculating excited states of Hamiltonians. WFQAE can calculate the $p$th excited state or up to the $p$th excited state depending on how we choose the weights and assign the circuit parameters. Compared to FQAE \cite{rahman2024feedback}, this algorithm has the advantage of only needing one optimization procedure to calculate the targeted eigenstate in contrast to FQAE, where all the lower energy eigenstates need to be calculated in an iterative manner. In addition, WFQAE does not require the calculation of inner products and overlap terms. We have shown the efficacy of WFQAE through an application in quantum chemistry, specifically through calculating the molecular excited states of the LiH molecule. 

This work can be further enhanced by investigating a suitable choice for the initial states. In addition, as explored in \cite{malla2024feedback}, designing a suitable cost Hamiltonian can enhance the algorithm's performance. 

Our algorithm advances quantum feedback-based methodologies, extending their applicability to a broader class of problems within the quantum computing landscape. This work highlights the capacity of quantum control theory to ignite the advancement of efficient quantum algorithms.

\appendix

This appendix explores extending the analysis for the case of multiple control inputs. Consider the model \eqref{model2} being modified as follows:
\begin{equation}
    \mathrm{i} |\dot{\Phi}(t)\rangle=(\hat H_d+\sum_{j=1}^d H_{c,j} \alpha^{(j)}(t))|\Phi(t)\rangle,
    \label{model3}
\end{equation}
where $\{\alpha^{(j)}(t)\}_{j=1, \dots,d}$ are the control inputs, with each corresponding to $H_{c,j}$. Considering similar Lyapunov function as \eqref{WL}, its derivative along the trajecories of model \eqref{model3} is given as
\begin{align}
\dot{V} & = \sum_{j=1}^d \sum_{q=0}^p w_q \bra*{\phi^{(q)}(t)} \mathrm{i} [H_{c,j},H_d] \ket*{\phi^{(q)}(t)} \alpha^{(j)} (t).
\label{vdot3}
\end{align}
We design $\alpha^{(j)} (t)$ such that $\dot{V}\leq0$:
\begin{equation} \label{controller2_m}
    \alpha^{(j)} (t)= - K_{g,j}   h \Big( \sum_{q=0}^p w_q \bra*{\phi^{(q)}(t)} \mathrm{i} [H_{c,j},H_d] \ket*{\phi^{(q)}(t)} \Big).
\end{equation}
By employing a similar Trotterization method as in Section~\ref{s2}, we get the following modification of \eqref{evolution2} :
\begin{align}
\ket*{\phi^{(q)}_l}  =  \prod_{k=1}^{l} \bigl(U_c(\alpha_k^{(1)}, \dots, \alpha_k^{(j)} )U_d \bigr) \ket*{\phi_0^{(q)}},
\label{evolution2_m}
\end{align}
where $U_c(\alpha_k^{(1)}, \dots, \alpha_k^{(j)} )= e^{-i \Delta t \sum_{j=1}^d \alpha^{(j)}(k \Delta t)  H_{c,j}}$, and the feedback law \eqref{controller2_dis} being modified as follows:
\begin{align} \label{controller2_dis_m}
    \alpha_{k+1} ^{(j)} &= - K_{g,j}   h \Big( \sum_{q=0}^p w_q \bra*{\phi^{(q)}_k} \mathrm{i} [H_{c,j},H_d] \ket*{\phi^{(q)}_k} \Big).       
\end{align}

\addtolength{\textheight}{0cm}   



\section*{Acknowledgment}
This work was supported by Independent Research Fund Denmark (DFF), project number 0136-00204B.

\bibliography{References.bib} 

\begin{thebibliography}{10}
\providecommand{\url}[1]{#1}
\csname url@samestyle\endcsname
\providecommand{\newblock}{\relax}
\providecommand{\bibinfo}[2]{#2}
\providecommand{\BIBentrySTDinterwordspacing}{\spaceskip=0pt\relax}
\providecommand{\BIBentryALTinterwordstretchfactor}{4}
\providecommand{\BIBentryALTinterwordspacing}{\spaceskip=\fontdimen2\font plus
\BIBentryALTinterwordstretchfactor\fontdimen3\font minus \fontdimen4\font\relax}
\providecommand{\BIBforeignlanguage}[2]{{%
\expandafter\ifx\csname l@#1\endcsname\relax
\typeout{** WARNING: IEEEtran.bst: No hyphenation pattern has been}%
\typeout{** loaded for the language `#1'. Using the pattern for}%
\typeout{** the default language instead.}%
\else
\language=\csname l@#1\endcsname
\fi
#2}}
\providecommand{\BIBdecl}{\relax}
\BIBdecl

\bibitem{cerezo2021variational}
M.~Cerezo, A.~Arrasmith, R.~Babbush, S.~C. Benjamin, S.~Endo, K.~Fujii, J.~R. McClean, K.~Mitarai, X.~Yuan, L.~Cincio \emph{et~al.}, ``Variational quantum algorithms,'' \emph{Nature Reviews Physics}, vol.~3, no.~9, pp. 625--644, 2021.

\bibitem{tang2021qubit}
H.~L. Tang, V.~Shkolnikov, G.~S. Barron, H.~R. Grimsley, N.~J. Mayhall, E.~Barnes, and S.~E. Economou, ``qubit-adapt-vqe: An adaptive algorithm for constructing hardware-efficient ans{\"a}tze on a quantum processor,'' \emph{PRX Quantum}, vol.~2, no.~2, p. 020310, 2021.

\bibitem{zhu2022adaptive}
L.~Zhu, H.~L. Tang, G.~S. Barron, F.~Calderon-Vargas, N.~J. Mayhall, E.~Barnes, and S.~E. Economou, ``Adaptive quantum approximate optimization algorithm for solving combinatorial problems on a quantum computer,'' \emph{Physical Review Research}, vol.~4, no.~3, p. 033029, 2022.

\bibitem{lavrijsen2020classical}
W.~Lavrijsen, A.~Tudor, J.~M{\"u}ller, C.~Iancu, and W.~De~Jong, ``Classical optimizers for noisy intermediate-scale quantum devices,'' in \emph{2020 IEEE international conference on quantum computing and engineering (QCE)}.\hskip 1em plus 0.5em minus 0.4em\relax IEEE, 2020, pp. 267--277.

\bibitem{magann2022feedback}
A.~B. Magann, K.~M. Rudinger, M.~D. Grace, and M.~Sarovar, ``Feedback-based quantum optimization,'' \emph{Physical Review Letters}, vol. 129, no.~25, p. 250502, 2022.

\bibitem{magann2022lyapunov}
------, ``Lyapunov-control-inspired strategies for quantum combinatorial optimization,'' \emph{Physical Review A}, vol. 106, no.~6, p. 062414, 2022.

\bibitem{DavidWakeham2021}
D.~Wakeham and J.~Ceroni, ``Feedback-based quantum optimization (falqon),'' \url{https://pennylane.ai/qml/demos/tutorial\_falqon/}, 05 2021, date Accessed: 2024-04-17.

\bibitem{larsen2023feedback}
J.~B. Larsen, M.~D. Grace, A.~D. Baczewski, and A.~B. Magann, ``Feedback-based quantum algorithm for ground state preparation of the fermi-hubbard model,'' \emph{arXiv preprint arXiv:2303.02917}, 2023.

\bibitem{magann2023randomized}
A.~B. Magann, S.~E. Economou, and C.~Arenz, ``Randomized adaptive quantum state preparation,'' \emph{arXiv preprint arXiv:2301.04201}, 2023.

\bibitem{motta2022emerging}
M.~Motta and J.~E. Rice, ``Emerging quantum computing algorithms for quantum chemistry,'' \emph{Wiley Interdisciplinary Reviews: Computational Molecular Science}, vol.~12, no.~3, p. e1580, 2022.

\bibitem{ura2023analysis}
K.~Ura, T.~Imoto, T.~Nikuni, S.~Kawabata, and Y.~Matsuzaki, ``Analysis of the shortest vector problems with quantum annealing to search the excited states,'' \emph{Japanese Journal of Applied Physics}, vol.~62, no.~SC, p. SC1090, 2023.

\bibitem{motta2020determining}
M.~Motta, C.~Sun, A.~T. Tan, M.~J. O’Rourke, E.~Ye, A.~J. Minnich, F.~G. Brandao, and G.~K.-L. Chan, ``Determining eigenstates and thermal states on a quantum computer using quantum imaginary time evolution,'' \emph{Nature Physics}, vol.~16, no.~2, pp. 205--210, 2020.

\bibitem{seki2021excited}
Y.~Seki, Y.~Matsuzaki, and S.~Kawabata, ``Excited state search using quantum annealing,'' \emph{Journal of the Physical Society of Japan}, vol.~90, no.~5, p. 054002, 2021.

\bibitem{bauman2020toward}
N.~P. Bauman, H.~Liu, E.~J. Bylaska, S.~Krishnamoorthy, G.~H. Low, C.~E. Granade, N.~Wiebe, N.~A. Baker, B.~Peng, M.~Roetteler \emph{et~al.}, ``Toward quantum computing for high-energy excited states in molecular systems: quantum phase estimations of core-level states,'' \emph{Journal of Chemical Theory and Computation}, vol.~17, no.~1, pp. 201--210, 2020.

\bibitem{wen2024full}
J.~Wen, Z.~Wang, C.~Chen, J.~Xiao, H.~Li, L.~Qian, Z.~Huang, H.~Fan, S.~Wei, and G.~Long, ``A full circuit-based quantum algorithm for excited-states in quantum chemistry,'' \emph{Quantum}, vol.~8, p. 1219, 2024.

\bibitem{cortes2022quantum}
C.~L. Cortes and S.~K. Gray, ``Quantum krylov subspace algorithms for ground-and excited-state energy estimation,'' \emph{Physical Review A}, vol. 105, no.~2, p. 022417, 2022.

\bibitem{nakanishi2019subspace}
K.~M. Nakanishi, K.~Mitarai, and K.~Fujii, ``Subspace-search variational quantum eigensolver for excited states,'' \emph{Physical Review Research}, vol.~1, no.~3, p. 033062, 2019.

\bibitem{higgott2019variational}
O.~Higgott, D.~Wang, and S.~Brierley, ``Variational quantum computation of excited states,'' \emph{Quantum}, vol.~3, p. 156, 2019.

\bibitem{mcclean2017hybrid}
J.~R. McClean, M.~E. Kimchi-Schwartz, J.~Carter, and W.~A. De~Jong, ``Hybrid quantum-classical hierarchy for mitigation of decoherence and determination of excited states,'' \emph{Physical Review A}, vol.~95, no.~4, p. 042308, 2017.

\bibitem{ryabinkin2018constrained}
I.~G. Ryabinkin, S.~N. Genin, and A.~F. Izmaylov, ``Constrained variational quantum eigensolver: Quantum computer search engine in the fock space,'' \emph{Journal of chemical theory and computation}, vol.~15, no.~1, pp. 249--255, 2018.

\bibitem{santagati2018witnessing}
R.~Santagati, J.~Wang, A.~A. Gentile, S.~Paesani, N.~Wiebe, J.~R. McClean, S.~Morley-Short, P.~J. Shadbolt, D.~Bonneau, J.~W. Silverstone \emph{et~al.}, ``Witnessing eigenstates for quantum simulation of hamiltonian spectra,'' \emph{Science advances}, vol.~4, no.~1, p. eaap9646, 2018.

\bibitem{rahman2024feedback}
S.~A. Rahman, {\"O}.~Karabacak, and R.~Wisniewski, ``Feedback-based quantum algorithm for excited states calculation,'' \emph{arXiv preprint arXiv:2404.04620}, 2024.

\bibitem{mackey2008deflation}
L.~Mackey, ``Deflation methods for sparse pca,'' \emph{Advances in neural information processing systems}, vol.~21, 2008.

\bibitem{magann2021digital}
A.~B. Magann, M.~D. Grace, H.~A. Rabitz, and M.~Sarovar, ``Digital quantum simulation of molecular dynamics and control,'' \emph{Physical Review Research}, vol.~3, no.~2, p. 023165, 2021.

\bibitem{cong2013survey}
S.~Cong and F.~Meng, ``A survey of quantum lyapunov control methods,'' \emph{The Scientific World Journal}, vol. 2013, 2013.

\bibitem{grivopoulos2003lyapunov}
S.~Grivopoulos and B.~Bamieh, ``Lyapunov-based control of quantum systems,'' in \emph{42nd IEEE International Conference on Decision and Control (IEEE Cat. No. 03CH37475)}, vol.~1.\hskip 1em plus 0.5em minus 0.4em\relax IEEE, 2003, pp. 434--438.

\bibitem{fedorov2022vqe}
D.~A. Fedorov, B.~Peng, N.~Govind, and Y.~Alexeev, ``Vqe method: a short survey and recent developments,'' \emph{Materials Theory}, vol.~6, no.~1, p.~2, 2022.

\bibitem{hempel2018quantum}
C.~Hempel, C.~Maier, J.~Romero, J.~McClean, T.~Monz, H.~Shen, P.~Jurcevic, B.~P. Lanyon, P.~Love, R.~Babbush \emph{et~al.}, ``Quantum chemistry calculations on a trapped-ion quantum simulator,'' \emph{Physical Review X}, vol.~8, no.~3, p. 031022, 2018.

\bibitem{zong2024determination}
Z.~Zong, S.~Huai, T.~Cai, W.~Jin, Z.~Zhan, Z.~Zhang, K.~Bu, L.~Sui, Y.~Fei, Y.~Zheng \emph{et~al.}, ``Determination of molecular energies via variational-based quantum imaginary time evolution in a superconducting qubit system,'' \emph{Science China Physics, Mechanics \& Astronomy}, vol.~67, no.~4, pp. 1--11, 2024.

\bibitem{malla2024feedback}
R.~K. Malla, H.~Sukeno, H.~Yu, T.-C. Wei, A.~Weichselbaum, and R.~M. Konik, ``Feedback-based quantum algorithm inspired by counterdiabatic driving,'' \emph{arXiv preprint arXiv:2401.15303}, 2024.

\end{thebibliography}

\end{document}